Limited surface mobility inhibits stable glass formation for 2-ethyl-1-hexanol

M. Tylinski[A], M. S. Beasley[A], Y. Z. Chua[B], C. Schick[B], M. D. Ediger[A*]

[A] Department of Chemistry, University of Wisconsin–Madison, Madison, Wisconsin 53706, USA

[B] Institute of Physics, University of Rostock, Albert-Einstein-Str. 23-24, 18051 Rostock, Germany and Competence Centre CALOR, Faculty of Interdisciplinary Research, University of Rostock, Albert-Einstein-Str. 25, 18051 Rostock, Germany

[*] Corresponding Author

*To be submitted to the Journal of Chemical Physics*

**Abstract:**

*Previous work has shown that vapor-deposition can prepare organic glasses with extremely high kinetic stabilities and other properties that would be expected from liquid-cooled glasses only after aging for thousands of years or more. However, recent reports have shown that some molecules form vapor-deposited glasses with only limited kinetic stability when prepared using conditions expected to yield a stable glass. In this work, we vapor deposit glasses of 2-ethyl-1-hexanol over a wide range of deposition rates and test several hypotheses for why this molecule does not form highly stable glasses under normal deposition conditions. The kinetic stability of 2-ethyl-1-hexanol glasses is found to be highly dependent on deposition rate. For deposition at $T_{substrate}$ = 0.90 $T_g$, the kinetic stability increases by 3 orders of magnitude (as measured by isothermal transformation times) when the deposition rate is decreased from 0.2 nm/s to 0.005 nm/s. We also find that, for the same preparation time, a vapor-deposited glass has much more kinetic stability than an aged liquid-cooled glass. Our results support the hypothesis that the formation of highly stable 2-ethyl-1-hexanol glasses is inhibited by limited surface mobility. We compare our deposition rate experiments to similar ones performed with ethylcyclohexane (which forms glasses of high kinetic stability); we estimate that the surface mobility of 2-ethyl-1-hexanol is more than 4 orders of magnitude less than that of ethylcyclohexane at 0.85 $T_g$.*

## Introduction:

There has been considerable effort in the last twenty years to understand mobility at the free surface of organic glasses, both polymeric and non-polymeric.[1,2] For polymers, high mobility at the free surface is often invoked to explain the decrease in the glass transition temperature ($T_g$) observed for thin (~10 nm) films of many different polymers.[3–8] High surface mobility has also been implicated in the unusual embedding behavior exhibited by nanoparticles at the surface of polymer films.[9] For low molecular weight glassformers, enhanced surface mobility has been linked to enhanced surface crystallization rates that may limit the utility of amorphous pharmaceuticals.[10–13] Enhanced surface mobility is also thought to play a key role in the preparation of vapor-deposited molecular glasses with high kinetic stability, known as stable glasses.[14,15] The properties of stable glasses resemble those expected for liquid-cooled glasses that have been aged for thousands of years or more.[16] Vapor-deposited glasses are also used in organic electronics and their distinctive features, in comparison to liquid-cooled glasses, are important for the function of these devices.[17,18] Our understanding of the phenomenon of enhanced surface mobility in both polymeric and non-polymeric glassformers has been aided by simulations[19–27] and by theoretical studies.[28–31]

Stable glasses are characterized by their high kinetic stability in comparison to liquid-cooled glasses.[16,32] The kinetic stability of a glass can be quantified by the temperature at which it begins to transform into the supercooled liquid upon heating. For stable glasses, the transformation temperature can be up to 8% higher than the conventional $T_g$.[33] In addition, the isothermal transformation of stable glass into the supercooled liquid can take up to $10^5$ times longer than the supercooled liquid structural relaxation time ($\tau_\alpha$).[34] Qualitatively, these indications of high kinetic stability are consistent with results expected for highly aged liquid-cooled glasses. The low molar volumes[35–37] and low enthalpies[16,38,39] exhibited by stable glasses also resemble properties expected for liquid-cooled glasses after extremely long aging times. Experiments,[14,16,40] simulations,[41,42] and theory[28] suggest that stable glasses are the result of high surface mobility that allows molecules to sample low energy configurations during deposition even though the temperature is below $T_g$. After continued deposition, surface molecules become part of the bulk. The low energy configurations found during deposition have very high barriers to rearrangement, giving rise to the kinetic stability of stable glasses. Configurations with similarly high barriers are inaccessible to liquid-cooled glasses on laboratory time scales because bulk mobility is extremely slow below $T_g$.

Although stable glasses have been prepared from more than 30 organic molecules, our recent report[34] found several molecules that failed to form stable glasses under similar deposition conditions. There are three hypotheses for why some molecules do not form highly stable glasses. First, it could be that amorphous packing arrangements with higher barriers to relaxation do not exist for these systems. In such a scenario, the configurations that would be found during deposition, even with a highly mobile surface, would have relaxation barriers similar to a liquid-cooled glass. Some work in the literature supports the view that the equilibrium temperature dependence of $\tau_\alpha$ becomes Arrhenius below $T_g$;[43–46] for liquids with this behavior, the barriers to relaxation would grow slowly or not at all as the temperature decreases. Second, for some molecules, the lowest energy configurations at the surface might not, upon further deposition, lead to bulk configurations with high barriers. It has been found that many vapor-deposited glasses have anisotropic packing and/or anisotropic molecular orientation,[17,18] suggesting that the preferred surface configurations are different from the bulk.[47] While anisotropic glasses can have high kinetic stabilities,[18] it might not be universally true that preferred surface configurations result in high barriers in the bulk. Third, it may be that molecules that do not form highly stable glasses have limited surface mobility, and thus deposition at the standard rate (~ 0.2 nm/s) does not allow surface molecules enough time to find the low energy configurations. All the molecules that did not form stable glasses in our previous work participate in intermolecular hydrogen bonding. Chen and coworkers have suggested that hydrogen bonding networks can inhibit surface mobility,[48] and without sufficient surface mobility, stable glass formation would not be expected. This third hypothesis would predict that lower deposition rates would enable stable glass formation.

In the present work, we characterize the kinetic stability of 2-ethyl-1-hexanol glasses prepared by vapor deposition using a wide range of deposition rates, an approach that has previously been used successfully to understand stable glass formation for other organic molecules.[16,38,40,49,50] For comparison, we also performed experiments on an aged liquid-cooled glass. We characterized the kinetic stability of the glasses using AC nanocalorimetry experiments. The onset temperature for transformation into the supercooled liquid was measured during temperature ramping experiments and we also determined the glass to liquid transformation time during isothermal annealing experiments. These experiments allow us to test the three hypotheses described above and to deduce why 2-ethyl-1-hexanol does not form a highly stable glass when using standard deposition conditions. In our previous publication, we found that glasses of 2-ethyl-1-hexanol had a transformation temperature of 1.04 $T_g$ and an isothermal transformation time of $10^2$ $\tau_\alpha$.[34] While these values indicate significant kinetic stability in comparison to a liquid-cooled glass, for other organic molecules the same deposition rate can produce glasses that have transformation temperatures up to 1.08 $T_g$ and isothermal transformation times up to $10^5$ $\tau_\alpha$.[33,34] If limited surface mobility inhibits the formation of highly stable glasses of 2-ethyl-1-hexanol, then lowering the deposition rate should result in a significant increase in stability.

We find that the kinetic stability of vapor-deposited glasses of 2-ethyl-1-hexanol approaches values typical of highly stable glasses when the deposition rate is decreased below 0.2 nm/s. We also observe that a vapor-deposited glass prepared at 0.95 $T_g$ has greater kinetic stability than a liquid-cooled glass that has been aged at 0.95 $T_g$ for a time equal to the deposition time. These findings are consistent with the hypothesis that the formation of stable glasses of 2-ethyl-1-hexanol is inhibited by limited surface mobility and allow us to rule out the other two hypotheses mentioned above. We estimate the surface mobility of 2-ethyl-1-hexanol by comparing our deposition rate results to those for ethylcyclohexane, which forms a highly stable glass when deposited at 0.2 nm/s.[40] We find that the surface mobility of 2-ethyl-1-hexanol is more than 4 orders of magnitude less than that of ethylcyclohexane at 0.85 $T_g$.

**Experimental:**

*Material*:

2-ethyl-1-hexanol with a purity $\geq$ 99.6% was purchased from Sigma Aldrich and used without further purification. The water content in the 2-ethyl-1-hexanol was 0.027% and was further reduced by pumping on the liquid prior to deposition. Throughout this paper, we make comparisons to the structural relaxation times ($\tau_\alpha$) for 2-ethyl-1-hexanol and to various data for ethylcyclohexane. The $\tau_\alpha$ data for 2-ethyl-1-hexanol and ethylcyclohexane supercooled liquids were obtained from literature sources.[51–54] We have previously applied a VTF fit to the low temperature regime of the $\tau_\alpha$ data and obtained $T_g$ ($\tau_\alpha$ = 100 s) values of 143.8 K and 100.9 K for 2-ethyl-1-hexanol and ethylcyclohexane respectively.[34]

*Apparatus:*

Glasses of 2-ethyl-1-hexanol were prepared by vapor-deposition onto an AC nanocalorimetry device housed inside an ultra-high vacuum chamber. AC nanocalorimetry experiments were then performed *in-situ*. The AC nanocalorimetry technique and our apparatus have been described in detail previously[55,56] Molecules for deposition are introduced into the vacuum chamber via a fine leak valve. The deposition is monitored by the ion gauge pressure and AC nanocalorimetry signal. The film thickness and deposition rate are calculated from the AC nanocalorimetry signal and a conversion factor determined from previously published *in-situ* ellipsometry experiments.[56] The base pressure in the chamber is typically below $10^{-7}$ Pa and ensures that there is no significant deposition of any contaminants during our experiments. By measuring the signal change on a calorimeter held at a typical deposition temperature, we estimate that films prepared using the lowest deposition rate are at least 99% 2-ethyl-1-hexanol. The AC nanocalorimeters are held in a copper housing that is attached to a liquid nitrogen reservoir. A heater and RTD attached to the copper housing allow for temperature control. The temperature of the sample is tentatively determined by measuring a resistor that is part of the AC nanocalorimeter active area. We have calibrated this resistor by measuring $\tau_\alpha$ of various supercooled liquids and our uncertainty in absolute temperature is +/- 1 K.[56] A more precise estimate of the temperature of each experiment is determined by measuring the 20 Hz $\tau_\alpha$ process of 2-ethyl-1-hexanol supercooled liquid after transformation of the glass and comparing to the literature value.[51] The temperature determined in this manner agreed with the previously described method within +/- 0.4 K.

For the nanocalorimetry experiment, a heater on the device membrane is driven by a 10 Hz AC current. The current produces heat at 20 Hz that results in a temperature oscillation of about 0.3 K at the sample. The temperature oscillation is detected by thermocouples patterned onto the device and the amplitude and phase of the voltage oscillation is measured by a lock-in amplifier. We perform a differential measurement between a device that has a sample on it and an empty device. We also perform a background subtraction to account for small differences between the devices. The resulting signal is proportional to the reversing heat capacity of the sample.[55] When comparing the heat capacities of samples of slightly different thicknesses, the signals were adjusted by comparing the data for the corresponding liquid-cooled glasses.

*Experiments:*

We performed temperature ramping and quasi-isothermal annealing experiments to characterize the kinetic stability of the vapor-deposited glasses. During temperature ramping experiments, we measure the reversing heat capacity while heating our samples at 5 K/min from 111 K (0.77 $T_g$) to 164 K (1.14 $T_g$). The temperature where the vapor-deposited glass begins to transform into the supercooled liquid ($T_{onset}$) is one measure of kinetic stability. The reversing heat capacity values of the vapor-deposited glasses are compared to a liquid-cooled glass prepared by cooling the sample from the supercooled liquid at 5 K/min. For quasi-isothermal annealing experiments, the glass is heated at 8 K/min to an annealing temperature near $T_g$. The annealing temperatures ranged from 146.5 K ($T_g$ + 2.7 K) to 141.2 K ($T_g$ – 2.6 K). The annealing temperatures were chosen to ensure that the transformation of glasses with low kinetic stability was slow enough for us to resolve and that the transformation of glasses with higher kinetic stability was completed before our liquid nitrogen supply was depleted. The transformation time for each experiment is normalized by $\tau_\alpha$ for the equilibrium supercooled liquid at the annealing temperature. It is known that some stable glasses transform via propagating fronts and the velocity of the transformation front is not quite proportional to $\tau_\alpha$;[56–58] if this applies to the transformation of 2-ethyl-1-hexanol glasses, the effect would be small compared to our error bars and so we have not attempted to correct the data. During isothermal annealing, we measured the reversing heat capacity to track the transformation of the glass into the supercooled liquid and to determine the transformation time, a second measure of kinetic stability. Films for temperature ramping experiments were approximately 120 nm thick. To prepare samples for quasi-isothermal annealing experiments, we deposited an approximately 250 nm thick film on top of a 120 nm layer of liquid-cooled glass. Previous work has shown that a stable glass in contact with either a supercooled liquid or a vacuum will initiate a transformation front with no induction time.[59] Thus, the bilayer sample geometry ensures that if the glass transforms via a front mechanism, there will be two fronts and both will start without any induction time.[56] The 0.90 $T_g$ glass deposited at 0.004 nm/s was only 170 nm thick (rather than 250 nm) due to the time limit associated with our liquid nitrogen supply. This thickness difference will shorten the transformation time by a factor of 1.5 or less, which is small compared to the changes observed in Figures 4 and 6. The positive error bar for this data point in Figures 4 and 6 accounts for this uncertainty.

**Results:**

*Temperature ramping experiments*

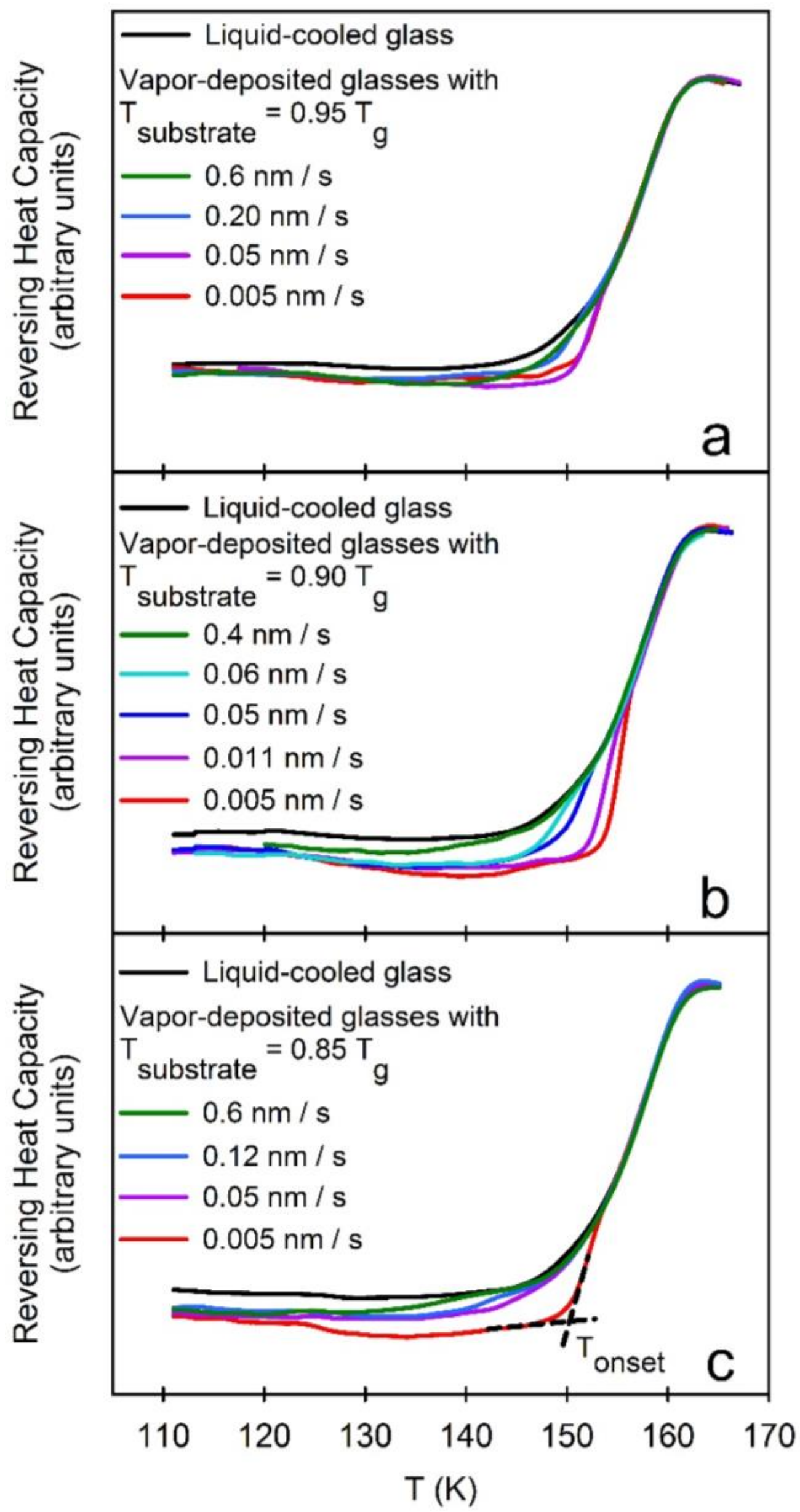

**Figure 1:** *Reversing heat capacity data from 20 Hz AC nanocalorimetry experiments on 2-ethyl-1-hexanol glasses, obtained during heating at 5 K/min. Panel **a**: Glasses deposited at $T_{substrate}$ = 0.95 $T_g$. Panel **b**: Glasses deposited at $T_{substrate}$ = 0.90 $T_g$. Panel **c**: Glasses deposited at $T_{substrate}$ = 0.85 $T_g$. The black dashed lines in panel c demonstrate how the onset temperature ($T_{onset}$) for the transformation of the as-deposited glass into the supercooled liquid is defined. For these experiments, the glasses are about 120 nm thick.*

Figure 1 shows the results of temperature ramping experiments on vapor-deposited glasses of 2-ethyl-1-hexanol that were used to determine the onset temperate ($T_{onset}$) for the transformation from as-deposited glass to supercooled liquid. The large step in the reversing heat capacity between 150 and 160 K occurs when the alpha relaxation process becomes fast enough to contribute to the 20 Hz reversing heat capacity. We determine the temperature at which the vapor-deposited glass begins to transform ($T_{onset}$) as indicated in Figure 1c. The $T_{onset}$ values are a measure of kinetic stability of the vapor-deposited glasses. Values greater than $T_g$ indicate that a vapor-deposited glass has greater kinetic stability than the liquid-cooled glass. The reversing heat capacity of the vapor-deposited glasses is lower than that of the liquid-cooled glass, as has been observed for many vapor-deposited organic glasses.[33]

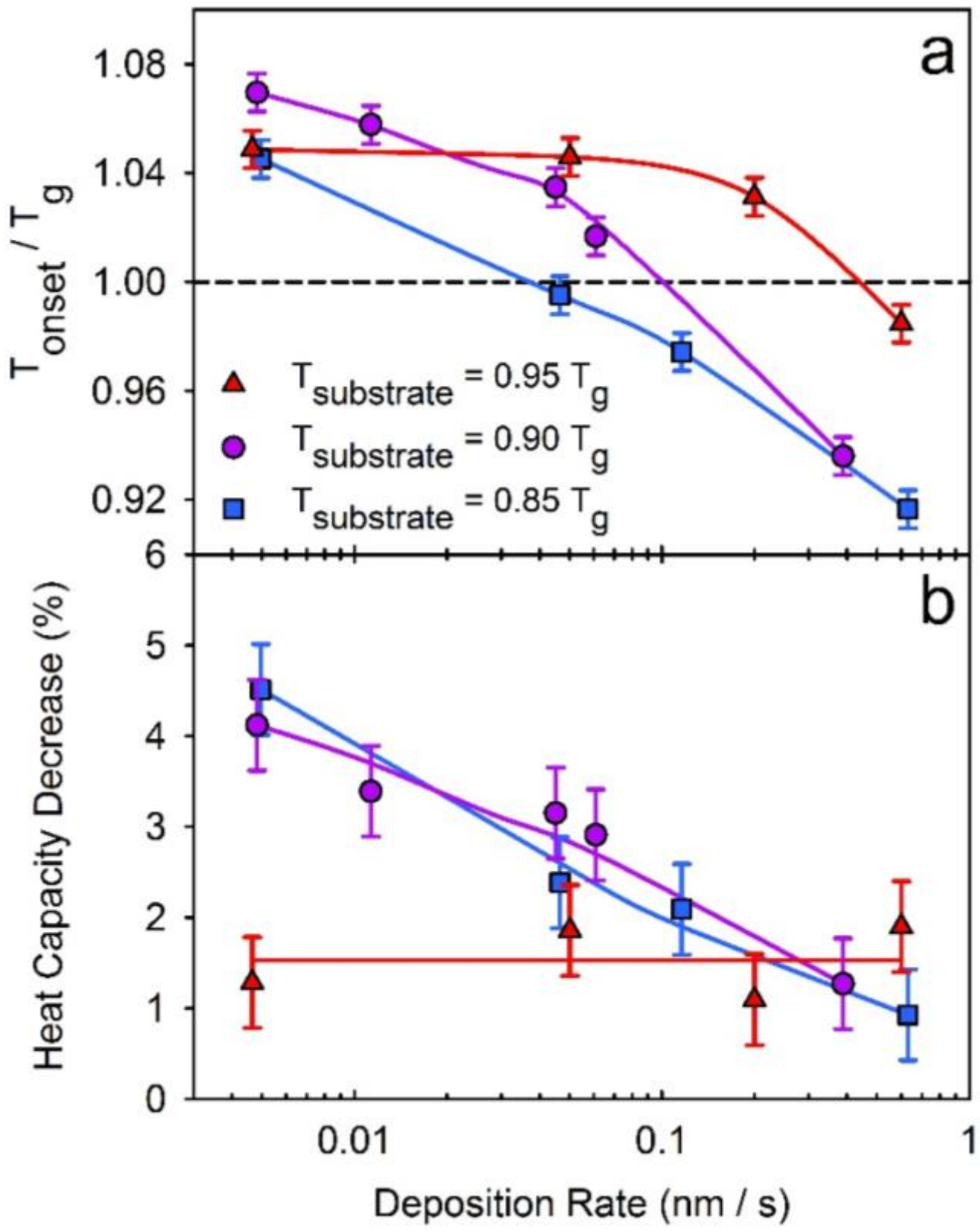

**Figure 2:** *Characteristics of vapor-deposited glasses of 2-ethyl-1-hexanol, as a function of deposition rate, from the temperature ramping experiments shown in Figure 1. Panel **a**: Onset temperature relative to $T_g$. Panel **b**: Heat capacity decrease of vapor-deposited glass compared to the liquid-cooled glass. The heat capacity decrease is determined at 0.95 $T_g$ (136.6 K). Lines are guides to the eye.*

Figure 2 summarizes how the deposition rate influences the $T_{onset}/T_g$ values (panel a) and the heat capacity decrease relative to the liquid-cooled glass (panel b) for vapor-deposited glasses of 2-ethyl-1-hexanol. The $T_{onset}/T_g$ values increase most significantly with decreasing deposition rate for the $T_{substrate}$ = 0.90 and 0.85 $T_g$ glasses. For the $T_{substrate}$ = 0.95 $T_g$ glasses, the $T_{onset}/T_g$ values increase and then reach a plateau for lower deposition rates. For the two lower substrate temperatures, the heat capacity of the as-deposited glass decreases significantly with lower deposition rates while the heat capacity for the $T_{substrate}$ = 0.95 $T_g$ glasses does not exhibit a trend with deposition rate. The error bars for the heat capacity values are large because we have used quite thin films for these experiments (120 nm). The thin films allow us to prepare samples with very low deposition rates while keeping the deposition time reasonable.

Figure 2 shows that the substrate temperature that yields the glass with the highest value of $T_{onset}/T_g$ changes as the deposition rate decreases. At rates near 0.2 nm/s the 0.95 $T_g$ glasses have the highest $T_{onset}$, which is consistent with our earlier work.[34] However, at lower deposition rates, the $T_{substrate}$ = 0.95 $T_g$ glasses reach a plateau in $T_{onset}/T_g$ and the $T_{substrate}$ = 0.90 $T_g$ glasses have higher values of $T_{onset}/T_g$. At high deposition rates, the $T_{substrate}$ = 0.85 $T_g$ glasses have very low values of $T_{onset}/T_g$, but at the lowest deposition rate the value increases to match that of the $T_{substrate}$ = 0.95 $T_g$ glass. The trend suggests that, at even lower deposition rates, the $T_{onset}/T_g$ values for the glasses deposited at 0.85 $T_g$ would surpass those of glasses deposited at 0.95 $T_g$. As we discuss below, all these results are consistent with the explanation that stable glass formation of 2-ethyl-1-hexanol deposited at 0.2 nm/s is inhibited by limited surface mobility.

*Quasi-isothermal transformation of vapor-deposited glasses*

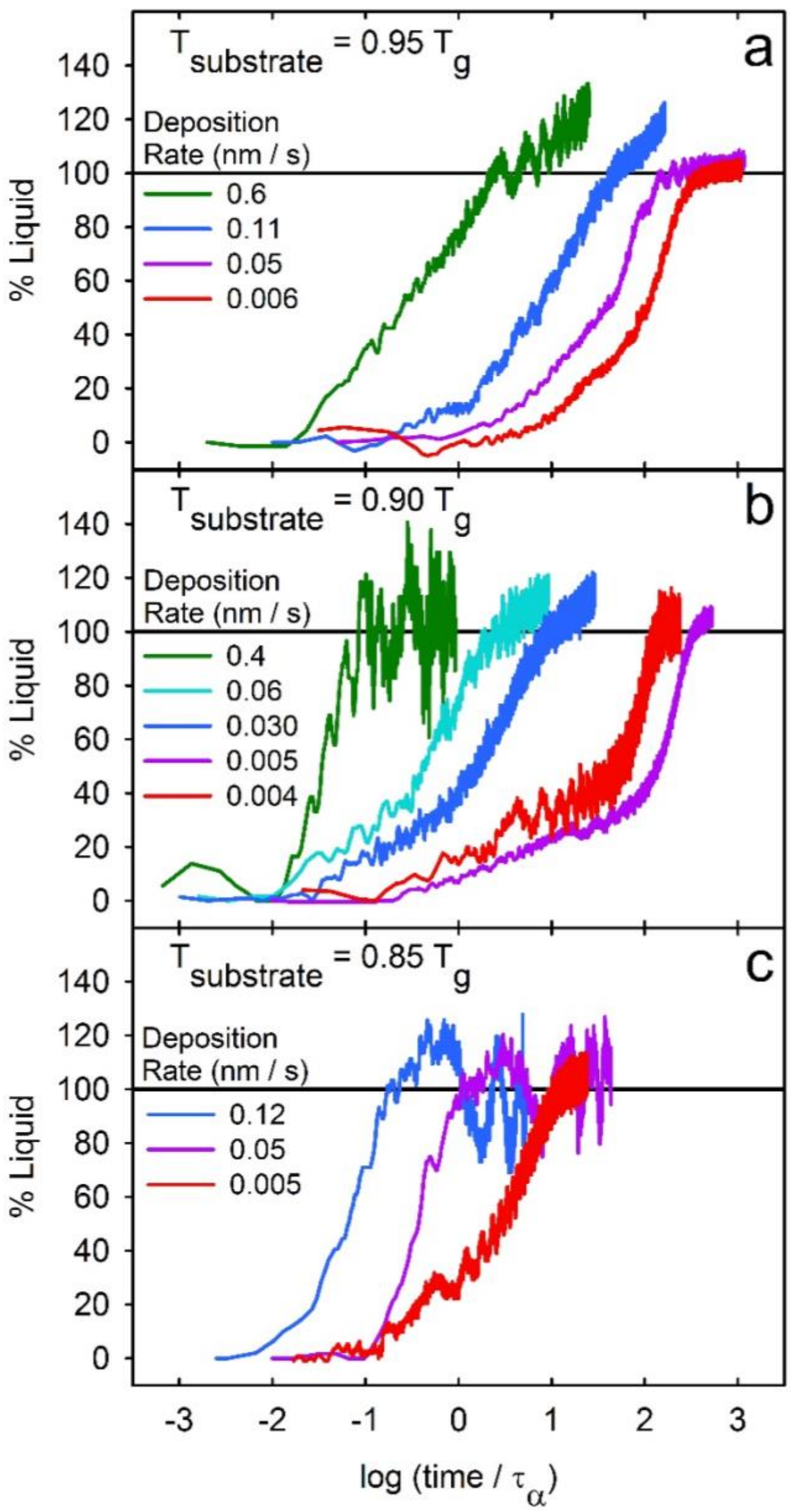

**Figure 3:** *Quasi-isothermal annealing experiments at $T_{anneal} \sim T_g$ for vapor-deposited glasses of 2-ethyl-1-hexanol prepared at different deposition rates. The "% Liquid" axis is determined from the reversing heat capacity increase from the as-deposited glass value to the supercooled liquid value. For the x-axis, the annealing time is normalized by the structural relaxation time of the supercooled liquid ($\tau_\alpha$) at the annealing temperature. Panels **a**, **b**, and **c** plot the transformation times for glasses deposited at $T_{substrate}$ = 0.95 $T_g$, 0.90 $T_g$, and 0.85 $T_g$ respectively. With one exception (see experimental section), the vapor-deposited glasses are 250 nm thick.*

We performed quasi-isothermal annealing experiments to measure how long it takes vapor-deposited glasses of 2-ethyl-1-hexanol to transform into the supercooled liquid when held near $T_g$. These transformation times are a second (and more precise) measure of kinetic stability. We measure the reversing heat capacity during the annealing experiment to track the transformation. The heat capacity increases from the as-deposited glass value until plateauing at the value of the supercooled liquid. In Figure 3 we have normalized this heat capacity increase and plot it as "% Liquid" to indicate what fraction of the sample is responding with the heat capacity of the supercooled liquid. We also have normalized the annealing time on the x-axis to account for annealing experiments performed at slightly different temperatures and to give a reference to the transformation times. We divide the annealing time by the structural relaxation time ($\tau_\alpha$) of the supercooled liquid at the annealing temperature. We expect that a liquid-cooled glass would transform in approximately 1 $\tau_\alpha$ and thus transformation times greater than $\tau_\alpha$ indicate that the vapor-deposited glass has greater kinetic stability than a liquid-cooled glass. The increase above the 100% level seen for some samples is due to instrument drift and has been previously observed in quasi-isothermal experiments.[33,40] It may be due to very slow deposition of molecules that are present in the chamber after the leak valve is closed at the end of the deposition. We present figures with linear time axes in the Supplemental Material that provide a clearer view of the end of the transformation and the determination of the transformation times (Figures S1, S2, and S3). The data in Figure 3 also show a low level oscillatory response that is an instrumental artifact.

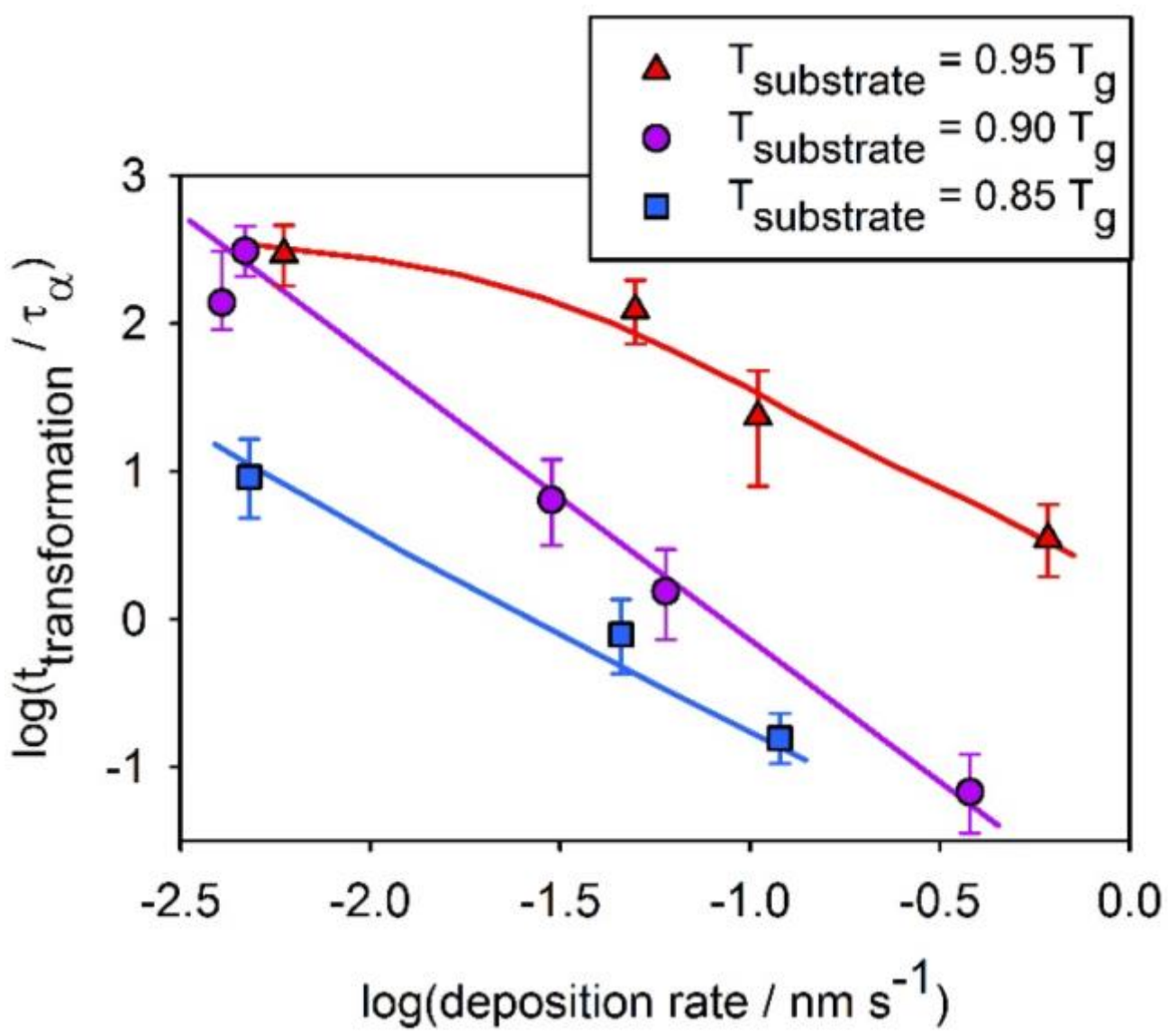

**Figure 4:** *Transformation times for vapor-deposited 2-ethyl-1-hexanol glasses during quasi-isothermal annealing experiments at $T_{anneal} \sim T_g$ as a function of deposition rate, as obtained from data in Figure 3. The lines are guides to the eye.*

We show the transformation times from the quasi-isothermal annealing experiments as a function of deposition rate in Figure 4. The transformation times of the vapor-deposited glasses generally increase with decreasing deposition rate. The $T_{substrate}$ = 0.95 $T_g$ glasses seem to approach a plateau in transformation time for low deposition rates; this is similar to the plateau in the $T_{onset}/T_g$ values shown for these glasses in Figure 2. For glasses deposited at 0.90 $T_g$, the transformation times increase by about 3 orders of magnitude as the deposition rate is decreased below the rate of 0.2 nm/s that has typically been used in our previous efforts to prepare stable glasses. At the lowest deposition rates the transformation times for 0.90 $T_g$ glasses meet the values for the 0.95 $T_g$ glasses and approach $\log(t_{transformation}/\tau_\alpha) = 3$. In our previous work we compiled isothermal transformation times for known stable glasses and found that they were all greater than $10^3$ $\tau_\alpha$.[34] That comparison utilized 590 nm thick films. Thin films of highly stable glasses are known to transform via a front mechanism. If we assume that the transformation time of the most stable glasses of 2-ethyl-1-hexanol scale with thickness, the longest transformation time in Figure 4 is equivalent to $10^{2.9}$ $\tau_\alpha$ for a 590 nm thick film, and very close to values associated with highly stable glasses.

*Aging experiments on a liquid-cooled glass*

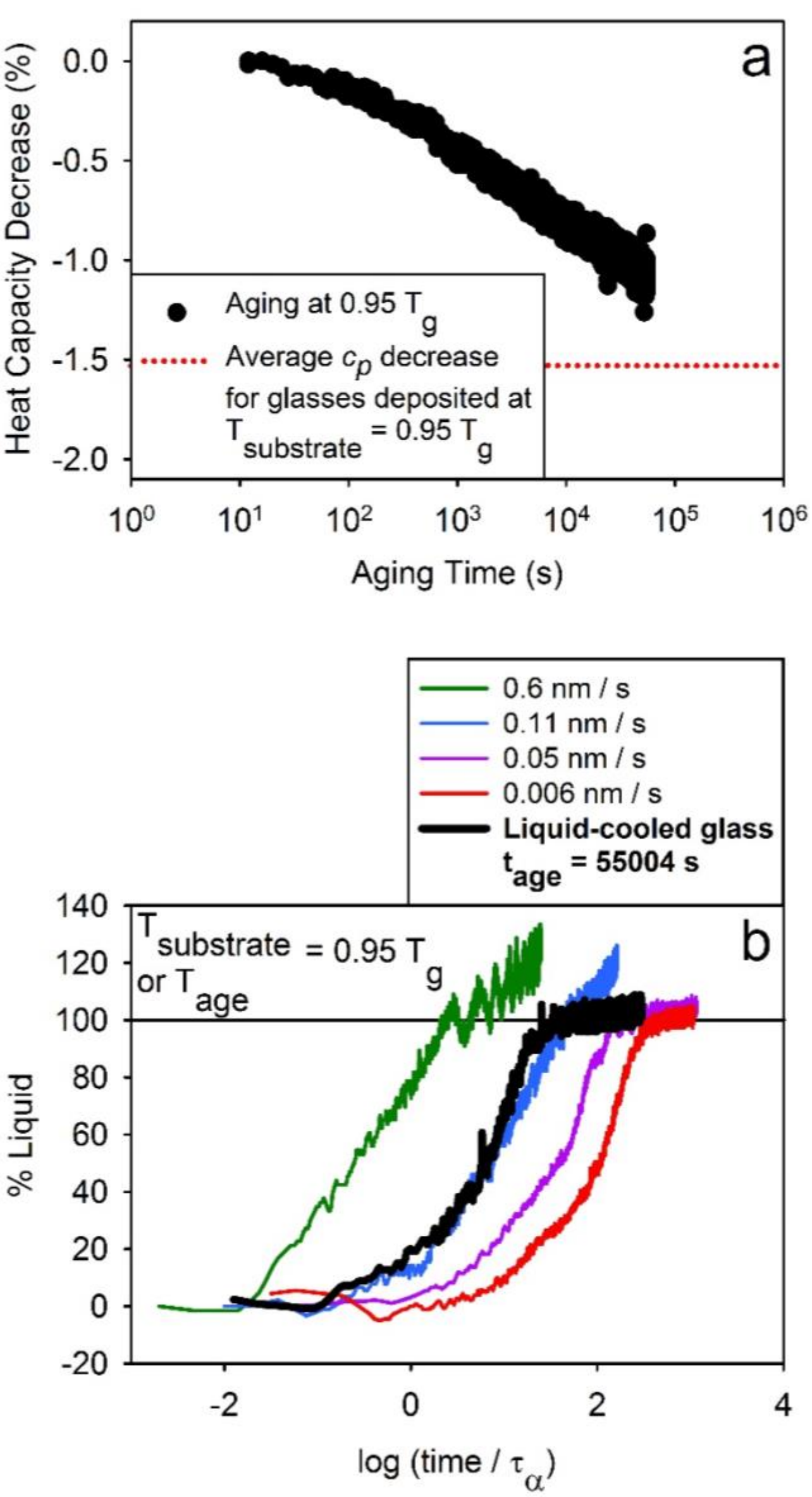

**Figure 5:** Aging experiments on a liquid-cooled glass of 2-ethyl-1-hexanol. *Panel **a**: Heat capacity decrease during aging at 0.95 $T_g$. The total aging time is equal to the time required for the longest deposition utilized to prepare the samples in Figure 3. For comparison, the red dashed line shows the average heat capacity decrease for glasses vapor-deposited at $T_{substrate}$ = 0.95 $T_g$. Panel **b**: Quasi-isothermal annealing transformation of the aged glass from panel a, in comparison to glasses vapor-deposited at $T_{substrate}$ = 0.95 $T_g$. The aged glass was about 370 nm thick.*

We aged a liquid-cooled glass to provide a comparison for the kinetic stability and heat capacity decrease of the vapor-deposited glasses. We prepared a glass by cooling the supercooled liquid at 5 K/min and aged it at 0.95 $T_g$ for 55,000 seconds. We chose this time to be equal to the longest time used for vapor deposition of the samples presented in this work. The decrease in reversing heat capacity during aging is shown in Figure 5a. The data in Figure 5a shows that this aging time was not sufficient to reach equilibrium (based upon the continuing decrease of the heat capacity). The heat capacity also did not reach the value of the heat capacity observed for the glasses vapor-deposited at $T_{substrate}$ = 0.95 $T_g$ (red dotted line).

We performed a quasi-isothermal annealing experiment on the aged glass and the results are overlaid in Figure 5b with those from the glasses vapor-deposited at $T_{substrate}$ = 0.95 $T_g$. The aged glass has a transformation time that is much less than the vapor-deposited sample prepared at 0.006 nm/s, despite similar preparation times. The aged glass transformation time is approximately the same as for the vapor-deposited glass prepared at 0.11 nm/s, whose preparation took only 2,300 seconds (24 times less than the aging time). These comparisons show that the kinetic stabilities of the vapor-deposited 2-ethyl-1-hexanol glasses are not simply a result of aging during the deposition process, and suggest that the kinetic stability of these vapor-deposited glass is controlled by surface mobility. We expand upon this argument in the discussion section.

**Discussion:**

The goal of this work is to investigate three hypotheses that offer an explanation for why 2-ethyl-1-hexanol does not form a highly stable glass when prepared at the conventional deposition rate of 0.2 nm/s. The first hypothesis is that 2-ethyl-1-hexanol does not possess amorphous configurations with barriers appreciably higher than those associated with the liquid cooled glass. In Figure 2a and Figure 4 we see that the kinetic stability of the vapor-deposited glasses increases greatly with decreasing deposition rate. For glasses prepared at $T_{substrate}$ = 0.90 $T_g$ and the lowest deposition rate, $T_{onset}$ reaches 1.07 $T_g$ and $\log(t_{transformation}/\tau_{\alpha})$ reaches $10^{2.5}$. These values of kinetic stability indicate that the configurations of these glasses have much higher barriers to relaxation than configurations typical of a liquid-cooled glass where $T_{onset} = T_g$ and $\log(t_{transformation}/\tau_{\alpha}) \sim 10^0$. The $T_{onset}$ and $\log(t_{transformation}/\tau_{\alpha})$ for 0.90 $T_g$ glasses are approaching values typical of highly stable glasses and the trend of the data suggest the kinetic stability is likely to increase further with lower deposition rates. Therefore, we reject the hypothesis that 2-ethyl-1-hexanol does not have amorphous configurations with high barriers.

The second hypothesis is that glasses of 2-ethyl-1-hexanol deposited at 0.2 nm/s have limited kinetic stability because the configurations preferred at the surface, upon further deposition, do not result in bulk packing with high kinetic stability. From this perspective, one would interpret the experimental observation that kinetic stability increases with decreasing deposition rate (Figure 4) as the result of physical aging that occurs during the deposition, since longer depositions are required at lower deposition rates. We performed the comparison between vapor-deposited glasses and an aged glass in Figure 5b to test this hypothesis. The transformation time of the liquid-cooled glass that aged for ~ 55,000 s was at least an order of magnitude less than for a vapor-deposited glass that took ~ 43,000 s to prepare (deposition rate = 0.006 nm/s). In addition, the glass vapor-deposited at 0.11 nm/s showed the same transformation time as the aged glass despite a preparation time that is shorter by a factor of 24. These observations in combination with the reasonably high kinetic stability achieved at the lowest deposition rates allow us to reject the second hypothesis.

The third hypothesis, that stable glass formation in 2-ethyl-1-hexanol is inhibited by limited surface mobility, is consistent with our experimental results as we discuss below. For the rest of the discussion we will interpret our results from this viewpoint. We will use our data to estimate the surface mobility of 2-ethyl-1-hexanol and make comparisons to molecules that form highly stable glasses using typical deposition rates.

*Kinetic stability as a function of deposition rate*

We observe that kinetic stability of the vapor-deposited 2-ethyl-1-hexanol glasses increases with decreasing deposition rate and that very low rates are needed to prepare glasses with high kinetic stability. We quantify kinetic stability using the onset temperature during temperature ramping experiments (Figure 2a) and the transformation time during quasi-isothermal annealing experiments (Figure 4). Many stable glasses display maximal onset temperatures of 1.05 – 1.08 $T_g$, and isothermal transformation times greater than $10^3$ $\tau_\alpha$.[33,34] By using deposition rates 2 orders of magnitude lower than a typical rate of 0.2 nm/s, we prepared 2-ethyl-1-hexanol glasses with $T_{onset}$ = 1.07 $T_g$ and $t_{transformation}$ = $10^{2.5}$ $\tau_\alpha$. The trend of the data suggests that for $T_{substrate}$ = 0.90 $T_g$ and 0.85 $T_g$, glasses prepared at lower deposition rates may have even greater kinetic stability.

The results of Figure 2a and Figure 4 shed light on the properties of amorphous configurations available at temperatures below $T_g$. The kinetic stability of the glasses deposited at 0.95 $T_g$ demonstrates that there are configurations at this temperature with higher barriers than those occupied at equilibrium at $T_g$. At high deposition rates, the 0.95 $T_g$ glasses have greater kinetic stability than 0.90 $T_g$ glasses, but the situation is reversed at the lowest rates. We interpret this to indicate that configurations with higher barriers are available at 0.90 $T_g$; however, since surface mobility decreases with decreasing temperature, surface molecules at 0.90 $T_g$ require more time to sample configurations that lead to high kinetic stability. It may be the case that there are configurations with even higher barriers at 0.85 $T_g$, but the surface mobility for 2-ethyl-1-hexanol at this temperature appears to be so slow that our lowest deposition rates do not allow nearly enough time for molecules to sample such configurations. In contrast, for molecules that more readily form stable glasses, surface mobility is fast enough that glasses with the highest kinetic stability are usually produced at $T_{substrate} \sim 0.85\ T_g$ using a rate of 0.2 nm/s.[33]

The results in Figures 2 and 4 at high deposition rates are also consistent with the hypothesis of limited surface mobility. Glasses formed at high deposition rates are quite unstable, transforming before $T_g$ or having transformation times shorter than $\tau_\alpha$, similar to a glass prepared by rapidly cooling a liquid. These results are to be expected if the surface mobility of 2-ethyl-1-hexanol is low enough that the molecules do not have enough time to sample configurations with even moderately high barriers at these high deposition rates.

*Surface vs. bulk relaxation times*

The kinetic stability of glasses prepared at $T_{substrate} = 0.95\ T_g$ appears to plateau at our lowest deposition rates, allowing us to estimate the time needed for surface relaxation at this temperature. The plateau indicates that the mobility of 2-ethyl-1-hexanol molecules is high enough that the surface has apparently equilibrated during deposition at the lowest rates. We can estimate a surface relaxation time in the following manner. At a deposition rate of ~ 0.03 nm/s, the onset temperature and transformation times of glasses deposited at 0.95 $T_g$ reaches or approaches a plateau. For this calculation, similar to previous work,[12,16,40] we estimate that the mobile layer is one monolayer thick (~ 0.6 nm for 2-ethyl-1-hexanol). At 0.03 nm/s, it takes 20 seconds to deposit one monolayer and we use this as an estimate of the surface relaxation time. This surface relaxation time is about 4 orders of magnitude faster than the equilibrium value of $\tau_\alpha$ for the bulk liquid, obtained by extrapolation. For ethylcyclohexane, a molecule that forms a highly stable glass when prepared at 0.2 nm/s, the surface relaxation time has been estimated to be 6-7 orders of magnitude faster than $\tau_\alpha$ at 0.95 $T_g$ using a similar analysis.[40] These comparisons support the hypothesis that 2-ethyl-1-hexanol has enhanced mobility at the surface in comparison to the bulk, but that this enhancement is not sufficient to enable the formation of highly stable glasses at 0.2 nm/s. (Experiments and simulations indicate that, depending on temperature, the mobile layer may be as thick as two or three monolayers,[14,20,60] rather than the single monolayer assumed above. These various possibilities have only a small effect on the order of magnitude comparisons that we make here.)

We can use the data in Figure 5b to develop a second comparison between the surface and bulk relaxation processes. The aged glass and the glass deposited at 0.11 nm/s have approximately the same transformation time. During the deposition of the 0.11 nm/s glass, each molecule had about 5.5 seconds as part of the mobile surface layer. Allowing each molecule 5.5 seconds as part of the surface layer results in the same kinetic stability as aging the liquid-cooled glass for 55,000 seconds. Thus, we estimate that the surface relaxation process is 4 orders of magnitude faster than the bulk process for 2-ethyl-1-hexanol at 0.95 $T_g$. It is reassuring that this estimate of the surface relaxation time is in good agreement with the completely independent estimate of the previous paragraph. From a similar analysis, Kearns and coworkers compared the fictive temperatures of indomethacin glasses that were aged or vapor-deposited at 0.94 $T_g$.[16] They estimated that the surface relaxation process for indomethacin was 7 orders of magnitude faster than the bulk aging process. From this comparison, we deduce that indomethacin has much faster surface relaxation than does 2-ethyl-1-hexanol at similar temperatures relative to $T_g$. This is consistent with the observation that indomethacin forms highly stable glasses when deposited at 0.2 nm/s, while 2-ethyl-1-hexanol does not. Yu and coworkers have used a surface grating decay method to measure surface diffusion on indomethacin glasses.[61] The surface diffusion coefficient at 0.95 $T_g$ is about 7 orders of magnitude faster than bulk diffusion. We do not know if surface diffusion is the most relevant measure of surface mobility for stable glass formation, but the grating decay experiments provide a comparison between bulk and surface dynamics that is broadly consistent with estimates obtained from vapor deposition experiments.

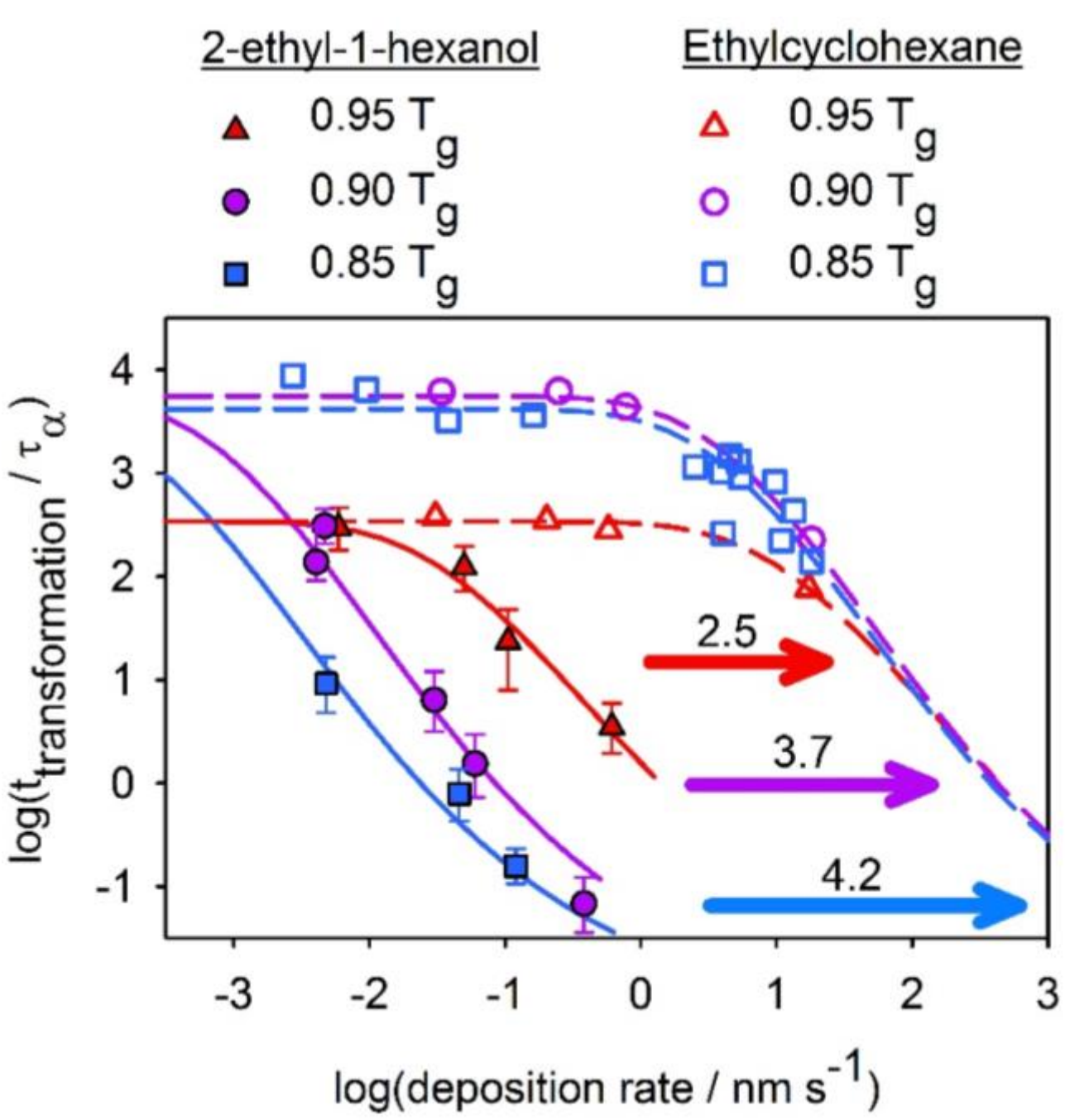

**Figure 6:** *Comparison of quasi-isothermal transformation times for vapor-deposited glasses of 2-ethyl-1-hexanol and ethylcyclohexane. The data for 2-ethyl-1-hexanol are the same as in Figure 4. The transformation times for ethylcyclohexane are slightly adjusted from the transformation times reported by Chua and coworkers,[40] as described in the text. The lines are stretched exponential fits to the data at each substrate temperature. Solid and dashed lines of the same color have the same plateau values at large and small deposition rates and are horizontally shifted by the number of decades above each colored arrow; this number is an estimate of the difference (in orders of magnitude) in surface relaxation times for the two molecules.*

*Estimation of surface relaxation times via comparison with ethylcyclohexane*

We estimate surface relaxation times for 2-ethyl-1-hexanol by comparing our deposition rate dependent transformation times to those of ethylcyclohexane glasses[40] as shown in Figure 6. The ethylcyclohexane experiments were performed on slightly thicker films (390 nm); assuming that these films transformed via a single propagation front, we have multiplied the transformation times from reference[40] by 0.32 to account for this difference. Chua and coworkers fit their transformation times to a Kohlrausch-Williams-Watts (KWW) function that plateaus at low deposition rates. A characteristic time ($\tau_{KWW}$) for each $T_{substrate}$ is a parameter obtained from the fit. We fit the data for both molecules together in order to obtain $\tau_{KWW}$ for these two systems. At each substrate temperature we apply the following fit to the transformation time data:

$$\log(t_{transformation}/\tau_\alpha) = A\left(1 - exp\left[-(t_{surface}/\tau_{KWW})^{\beta}\right]\right) + B \qquad (1)$$

The sum A+B gives the value of $\log(t_{transformation}/\tau_\alpha)$ in the limit of large $t_{surface}$ or low deposition rate. The amount of time on the surface is given by: $t_{surface}$ = 0.6 nm / deposition rate (nm/s). Following Chua and coworkers, the value of $\beta$ was fixed to 0.35. We assume the values of A and B at a given $T_{substrate}$ are the same for both molecules (see following paragraph). After preliminary fits in which B was allowed to vary with $T_{substrate}$, we decided to set B = -2.3 for all $T_{substrate}$ for simplicity; the effect of this on the $\tau_{surface}$ values in Figure 7 is within the error bars. To fit the 2-ethyl-1-hexanol and ethylcyclohexane data together, we shifted the 2-ethyl-1-hexanol $t_{surface}$ values for a given $T_{substrate}$ by a common factor, choosing the factor of maximize the quality of the fit to the combined data sets. The colored arrows in Figure 6 represent the size of the shift (in orders of magnitude) obtained from the fitting procedure and also give the number of orders of magnitude difference between $\tau_{KWW}$ for 2-ethyl-1-hexanol and ethylcyclohexane. As done by Chua and coworkers, we calculate the surface relaxation time as $\tau_{surface}$ = 10 $\tau_{KWW}$; this time represents an estimate of the minimum time on the surface needed to sample configurations that correspond to the highest observed kinetic stability. The $\tau_{surface}$ values are plotted in Figure 7 along with $\tau_{\alpha}$ data for both molecules. The $\tau_{surface}$ values for ethylcyclohexane at $T_{substrate} < 0.85\ T_g$ were obtained in a similar manner, but there was no 2-ethyl-1-hexanol data included for these substrate temperatures. The $\tau_{surface}$ values for ethylcyclohexane obtained here are in good agreement with those obtained in reference [40].

Our comparison above utilized a few assumptions and we suggest that these assumptions are reasonable. First, we assumed the ethylcyclohexane films transformed by fronts and adjusted the transformation times accordingly. The results of stable glasses formed from a number of different molecules have found that stable glasses that are 400 nm or thinner will transform via a front mechanism. Then, we assumed in our fitting that the low deposition rate limit of $\log(t_{transformation}/\tau_{\alpha})$ is the same for 2-ethyl-1-hexanol and ethylcyclohexane for a given value of $T_{substrate}/T_g$. We see in Figure 6 that for $T_{substrate} = 0.95\ T_g$, the data suggest that this is indeed the case. We also expect the molecules to have similar $\log(t_{transformation}/\tau_{\alpha})$ plateaus based on their similar supercooled liquid dynamics. Both the $\log(t_{transformation}/\tau_{\alpha})$ plateau value and $\tau_{\alpha}$ are controlled by the barrier heights for the lowest energy configurations at a given temperature. In Figure 7 we see that the VTF predictions of $\tau_{\alpha}$ below $T_g$ for 2-ethyl-1-hexanol and ethylcyclohexane are very similar. This implies that when normalized by $T_g$, the extrapolated barrier heights for these two liquids are the same at low temperatures. Thus, at a given $T_{substrate}/T_g$ and in the limit of low deposition rate, the low energy amorphous configurations for the two molecules should have similarly sized barriers relative to $T_g$ and we expect the transformation times of the two glasses to be similar. While the assumptions described above and the choice of a particular fitting function (eq. 1) have little impact on the value of $\tau_{surface}$ at 0.95 Tg, the possible error is larger at lower temperatures and not included in the error bars shown in Figure 7.

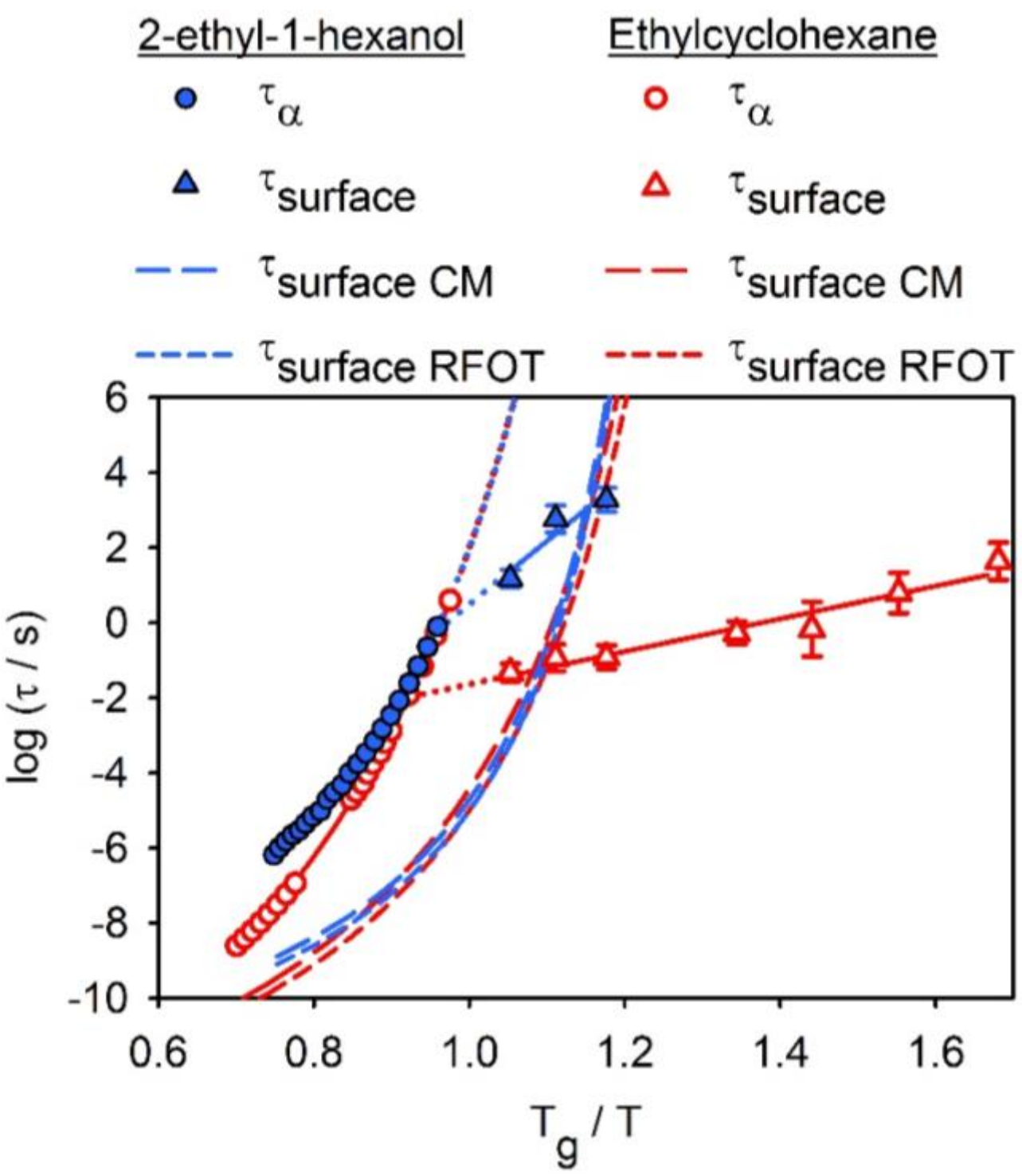

**Figure 7:** *Relaxation times $\tau_\alpha$ and $\tau_{surface}$ for 2-ethyl-1-hexanol and ethylcyclohexane. The $\tau_{surface}$ values are obtained by fitting the data shown in Figure 6 and data from reference [40] using the procedure described in the text. The $\tau_\alpha$ values for 2-ethyl-1-hexanol are from dielectric spectroscopy measurements in reference [51]. The $\tau_\alpha$ values for ethylcyclohexane are from dielectric spectroscopy measurements in references [52] and [53], and the calculation of $\tau_\alpha$ in reference [52] using viscosity measurements in reference [54]. Predictions of $\tau_{surface}$ from the Random First Order Transition (RFOT) theory[21] and the Coupling Model[23] (CM) are plotted as dashed lines. VTF fits are plotted with the $\tau_\alpha$ data. Arrhenius fits to the $\tau_{surface}$ data are shown. Dotted lines are used to indicate extrapolation of a fitting function beyond the range of experimental data.*

Figure 7 shows that the enhancement of surface mobility is less significant for 2-ethyl-1-hexanol than for ethylcyclohexane. At 0.95 $T_g$ ($T/T_g = 1.05$), $\tau_{surface}$ is about 4 orders of magnitude smaller than the extrapolated $\tau_\alpha$, consistent with the comparisons made earlier in the discussion section. As expected from Figure 6, we observe that $\tau_{surface}$ is longer for 2-ethyl-1-hexanol than for ethylcyclohexane and has a stronger temperature dependence. Work by Chen and coworkers[48] suggests that the ability to form intermolecular hydrogen bonds may limit the surface mobility and this would be consistent with the slower surface relaxation of 2-ethyl-1-hexanol.

In Figure 7, we also plot predictions of $\tau_{surface}$ from the Random First Order Transition (RFOT) theory[21] and the Coupling Model (CM).[23] For the RFOT calculation, we use $\tau_{surface} = (\tau_0\ \tau_\alpha)^{0.5}$ and a $\tau_0$ value of 1 ps. The CM predicts $\tau_{surface} = (t_c)^{1-\beta}(\tau_\alpha)^\beta$ where $t_c$ = 2 ps and $\beta$ is the stretching exponent from a KWW fit to the alpha relaxation; we used $\beta$ values of 0.51 for 2-ethyl-1-hexanol and 0.53 for ethylcyclohexane as obtained from dielectric spectroscopy experiments.[53,62] The predictions of the RFOT and CM approaches are quite similar and both predict a very different temperature dependence than what is obtained experimentally. Also, these calculations do not capture the significant difference between $\tau_{surface}$ for 2-ethyl-1-hexanol and ethylcyclohexane. We emphasize that the $\tau_{surface}$ value shown in Figure 7 are not direct measurements of this quantity but rather an inference based upon the stability of vapor-deposited glasses. It is possible that more direct measurements of surface relaxation for these compounds, such as the surface diffusion coefficients obtained from the grating decay method, would show behavior more similar to the predictions of references [28] and [30]. There has also been interest in assessing whether the beta relaxation is closely related to the surface relaxation time;[30] we make that comparison for these two molecules in Figure S4 of the Supplemental Material. For ethylcyclohexane, there is a significant difference between $\tau_{surface}$ and the Johari Goldstein beta relaxation time.

The apparent intersection of $\tau_{surface}$ and $\tau_\alpha$ at temperatures near 1.1 $T_g$ is an interesting phenomenon that presents a challenge to our understanding of enhanced surface mobility. Similar behavior has been previously observed for polymers and recently for low molecular weight glassformers.[2,5,9,40,63–66] From our perspective, this intersection (by extrapolation) at temperatures ≤ 1.1 $T_g$ (or $\tau_\alpha \geq 10^{-2}$ s) is surprising based upon the following understanding of supercooled liquid behavior. The $\tau_\alpha$ values of fragile supercooled liquids becomes very large with decreasing temperature because the alpha relaxation requires the cooperative motion of multiple molecules.[67] The number of cooperating molecules increases with decreasing temperature,[68] and thus $\tau_\alpha$ exhibits a super-Arrhenius temperature dependence. Molecules at the surface are expected to relax faster than those in the bulk because they have fewer neighbors to inhibit their motion.[28,29] Supercooled liquids commonly start showing evidence of cooperative dynamics for $\tau_\alpha > 10^{-11}$ s;[68–70] values much less than the intersection points observed in Figure 7. Based on this, we would expect $\tau_{surface}$ to remain faster than $\tau_\alpha$ up to much higher temperatures than is indicated by the data. Theoretical treatments of surface mobility also do not predict an intersection of $\tau_{surface}$ and $\tau_\alpha$ at any temperature.[28–30] The predictions of RFOT and CM theories in Figure 7 illustrate this behavior. Until recently, the evidence for the intersection of surface and bulk relaxation processes has largely come from experiments with polymers.[2,5,9,63,65,66] Studies of surface and bulk diffusion coefficients of low molecular weight glassformers do not show any indication of such an intersection.[10,12,13,61] It has been suggested that the intersection of the surface and bulk relaxation processes might be due to chain connectivity effects in polymers.[5,29] However, it appears that the intersection of surface and bulk processes also occurs for small molecules based on the $\tau_{surface}$ data in Figure 7 for 2-ethyl-1-hexanol and ethylcylohexane,[40] and based on thin film experiments with N,N'-Bis(3-methylphenyl)-N,N'-diphenylbenzidine (TPD).[64] While we do not have a way to reconcile all these results, it seems important to more fully understand what is being measured by the various experiments. It would be useful to perform surface diffusion measurements and deposition rate dependent vapor deposition on the same low molecular weight glassformer. Such experiments would be expected to contribute, although indirectly, to our understanding of dynamics in thin polymer films.

**Conclusion:**

We prepared vapor-deposited glasses of 2-ethyl-1-hexanol using a wide range of deposition rates and assessed the kinetic stability of the glasses using AC nanocalorimetry. At deposition rates typically used to prepare highly stable glasses (0.2 nm/s), 2-ethyl-1-hexanol forms glasses with moderate kinetic stability. We observed that the kinetic stability increases significantly when the deposition rate is decreased, consistent with the hypothesis that the surface mobility for this molecule is limited in comparison to those that form highly stable glasses when deposited at 0.2 nm/s. We performed experiments with an aged glass to confirm that the increase in kinetic stability is indeed a result of surface relaxation and is not due to bulk aging during the long deposition period. We used our results to estimate that the surface relaxation process in 2-ethyl-1-hexanol is faster than bulk relaxation, but is slower than the surface relaxation of molecules like ethylcylohexane and indomethacin that readily form highly stable glasses. We estimated $\tau_{surface}$ values for 2-ethyl-1-hexanol and compared these to results for ethylcyclohexane and to theoretical predictions. $\tau_{surface}$ values for both 2-ethyl-1-hexanol and ethylcyclohexane show an apparent intersection with $\tau_{\alpha}$ near 1.1 $T_g$ that is not predicted by the RFOT or Coupling Model. This feature, which is similar to reports for polymeric glassformers, deserves further investigation.

This work on 2-ethyl-1-hexanol was motivated by a previous study[34] comparing the properties of vapor-deposited glasses of six alcohols and we briefly discuss these previous results in light of our new findings. The majority of organic molecules tested thus far form highly stable glasses when vapor deposited at 0.2 nm/s at an appropriate substrate temperature. For some of these molecules, there is strong evidence that this occurs as a result of high surface mobility.[1,10,12] It is reasonable to assume that this is generally true for all molecules that can form highly stable glasses, including benzyl alcohol from reference 34. Ethanol forms a glass with slightly lower kinetic stability than does 2-ethyl-1-hexanol at 0.2 nm/s and so we suspect that slower deposition will have a similarly strong impact on the stability of ethanol glasses. Finally, when three of the alcohols studied in reference 34 (propanol, propylene glycol, and ethylene glycol) were deposited at 0.2 nm/s, they formed glasses with no increase of stability relative to liquid-cooled glasses. For these molecules, it may be the case that slower deposition would result in more stable glasses, in accord with the results for 2-ethyl-1-hexanol, although it is possible that this would not occur in the range of deposition rates that we can access experimentally. It is also possible that for this class of molecules a factor other than surface mobility is limiting stable glass formation.

**Acknowledgements:** We thank the U.S. National Science Foundation (CHE-1265737 and CHE-1564663) for support of this work. We thank Karen Jones for determining the water content in 2-ethyl-1-hexanol.

**Supplemental Material:** Supplemental material is available for this work.

**Supplemental Material for:**

Limited surface mobility inhibits stable glass formation for 2-ethyl-1-hexanol

M. Tylinski[A], M. S. Beasley[A], Y. Z. Chua[B], C. Schick[B], M. D. Ediger[A*]

[A] Department of Chemistry, University of Wisconsin–Madison, Madison, Wisconsin 53706, USA

[B] Institute of Physics, University of Rostock, Albert-Einstein-Str. 23-24, 18051 Rostock, Germany and Competence Centre CALOR, Faculty of Interdisciplinary Research, University of Rostock, Albert-Einstein-Str. 25, 18051 Rostock, Germany

[*] Corresponding Author

**Quasi-isothermal annealing experiments:**

In Figures S1 – S3 we replot of the quasi-isothermal annealing experiments shown in Figure 3 with linear time axes in order to clarify the transformation times for these experiments. We have added dashed lines to demonstrate the determination of the transformation time.

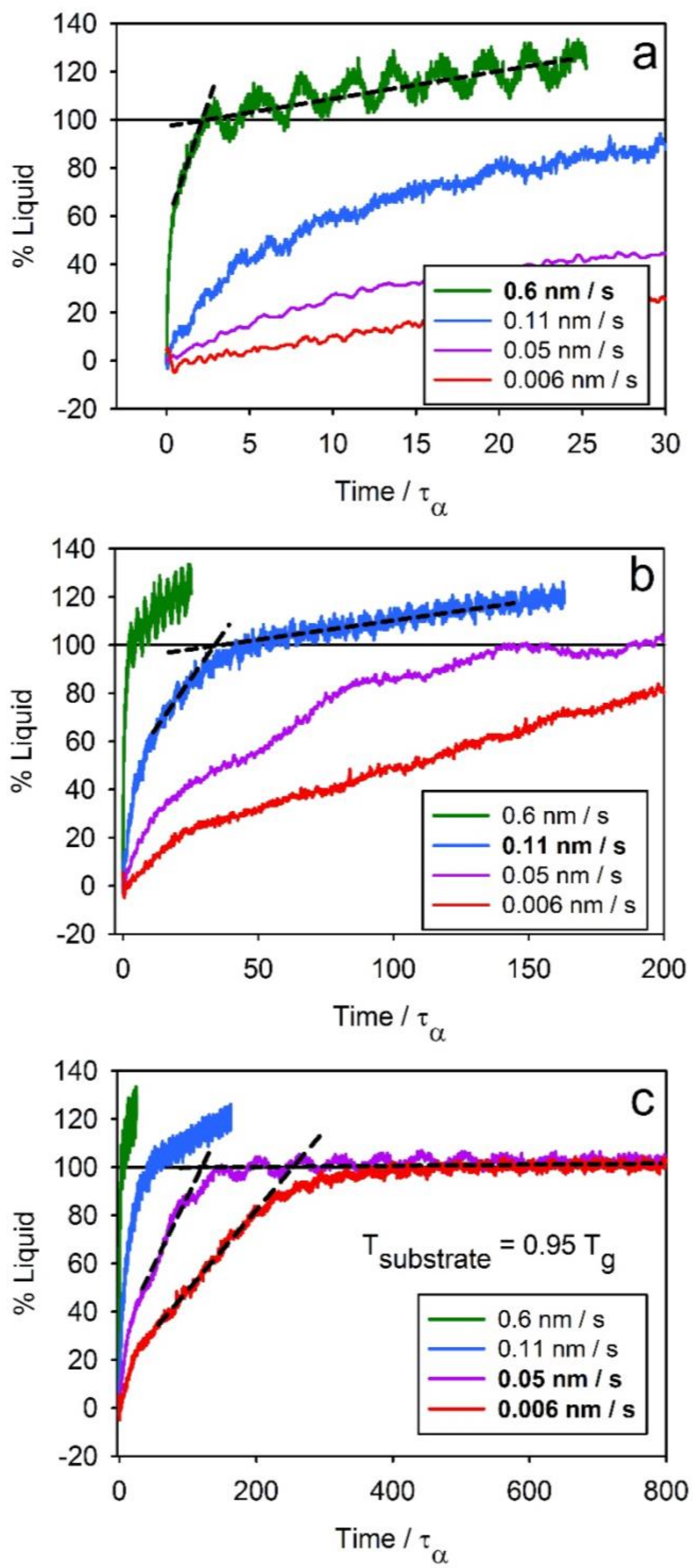

**Figure S1:** *Quasi-isothermal transformations of $T_{substrate}$ = 0.95 $T_g$ glasses with linear time axes.*

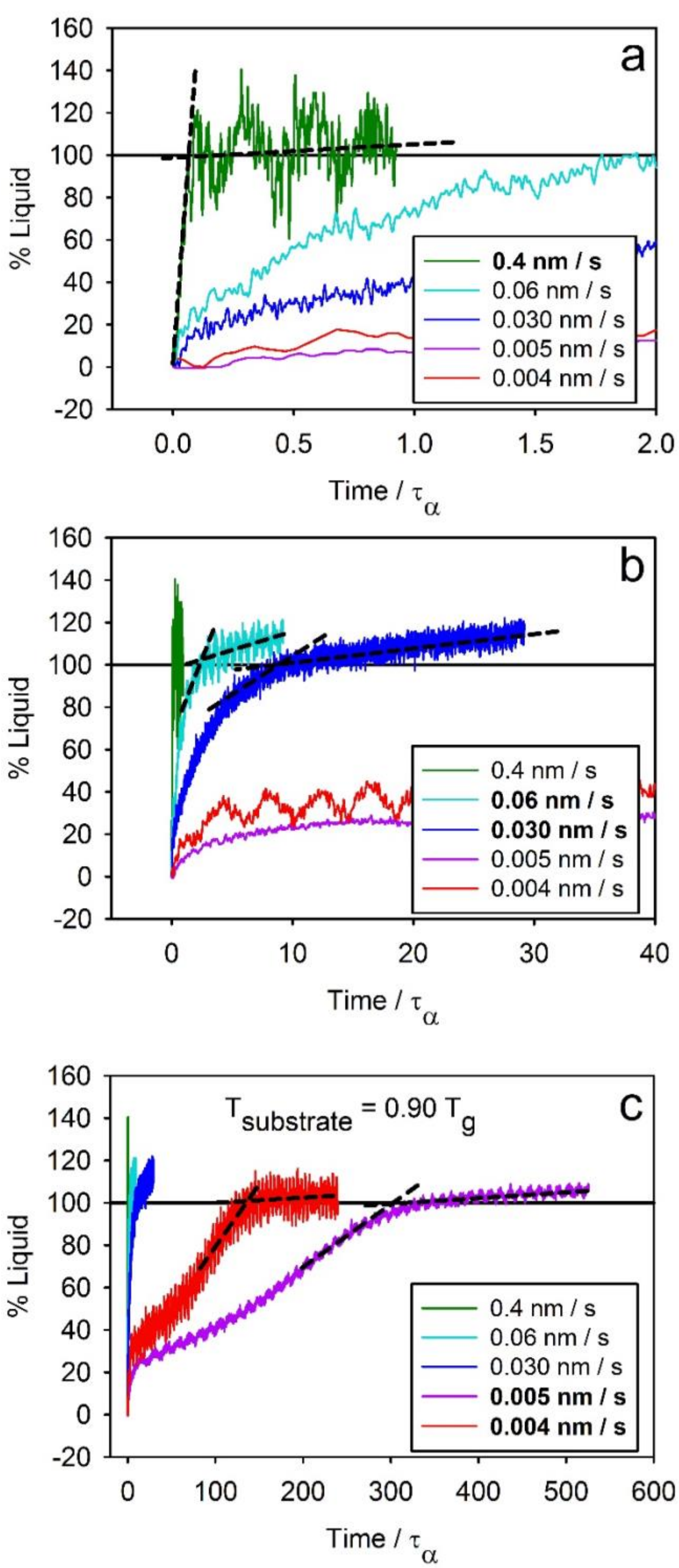

**Figure S2:** *Quasi-isothermal transformations of $T_{substrate}$ = 0.90 $T_g$ glasses with linear time axes. The film deposited at 0.004 nm/s was 170 nm thick, compared to 250 nm for the other samples. If the transformation time scales with thickness, the 0.004 nm/s film would have a similar transformation time compared to the 0.005 nm/s sample when adjusted for thickness.*

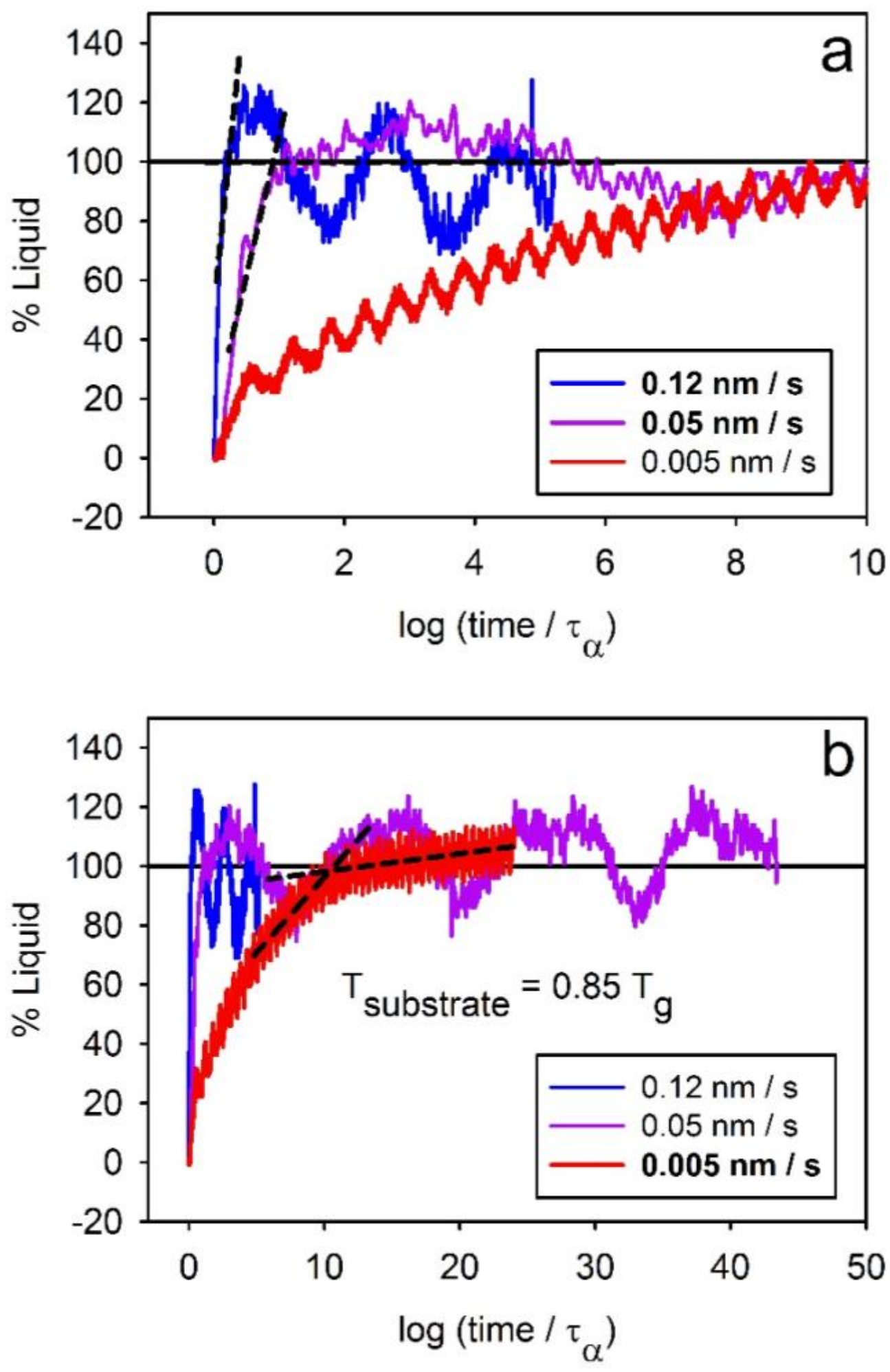

**Figure S3:** *Quasi-isothermal transformations of $T_{substrate}$ = 0.85 $T_g$ glasses with linear time axes.*

**Comparison of $\tau_{surface}$, $\tau_{\alpha}$, and $\tau_{\beta}$ relaxation times:**

In Figure S4 we plot the $\tau_{surface}$ and $\tau_{\alpha}$ values for 2-ethyl-1-hexanol and ethylcyclohexane (as in Figure 7) with the addition of experimentally determined and $\tau_{\beta}$ values. We have included the comparison between $\tau_{surface}$ and $\tau_{\beta}$ because it has been suggested that the beta relaxation process and the surface relaxation process are related and thus might be expected to have similar relaxation times.[1,2] The Coupling Model in particular predicts that the Johari Goldstein beta relaxation time will be approximately the same as the surface relaxation time ($\tau_{\beta\,JG} \approx \tau_{surface}$).[1] For 2-ethyl-1-hexanol the observed $\tau_{\beta}$ values do not agree with the $\tau_{surface}$ values and for ethylcyclohexane the calculated $\tau_{\beta\,JG}$ values do not agree with the $\tau_{surface}$ values.

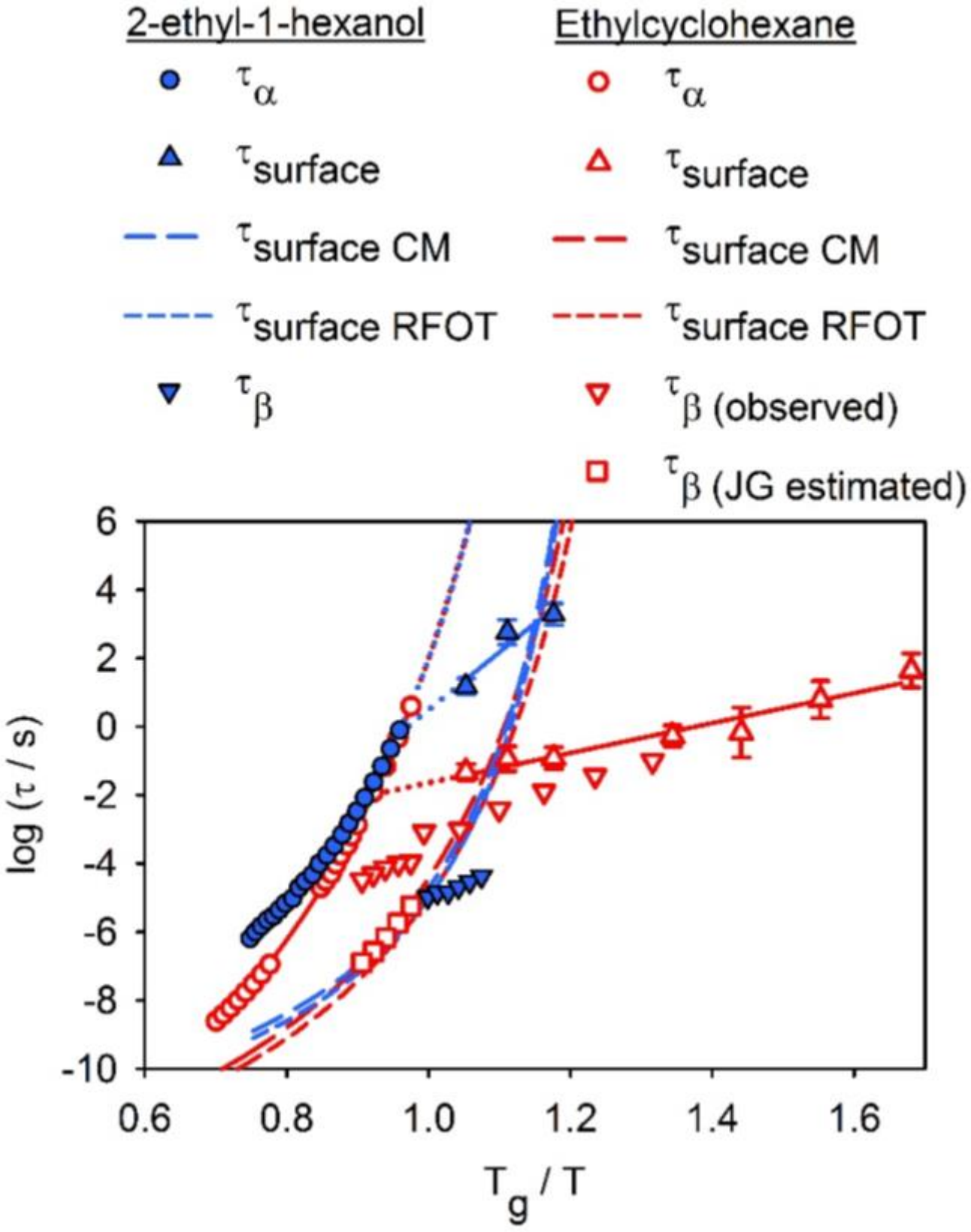

**Figure S4:** *Relaxation plot showing $\tau_\alpha$, $\tau_{surface}$, and $\tau_\beta$ values for 2-ethyl-1-hexanol and ethylcyclohexane. The $\tau_\alpha$ and $\tau_\beta$ values for 2-ethyl-1-hexanol are from dielectric spectroscopy measurements in reference* [3]*.The $\tau_{surface}$ values for 2-ethyl-1-hexanol are obtained by the comparison with ethylcyclohexane data shown in Figure 6. The $\tau_\alpha$ values for ethylcyclohexane are from dielectric spectroscopy measurements in references* [4] *and* [5]*, and the calculation of $\tau_\alpha$ in reference* [4] *using viscosity measurements in reference* [6]*. The $\tau_{\beta(observed)}$ values for ethylcyclohexane are from dielectric spectroscopy measurements in references* [4] *and* [5]*. The authors of those papers have argued that the observed beta relaxation is not the Johari-Goldstein (JG) beta relaxation. The $\tau_{\beta(JG\ estimated)}$ values for ethylcyclohexane were calculated by the authors in reference* [4] *based on the Coupling Model. The ethylcyclohexane $\tau_{surface}$ values were obtained by the fitting procedure described in the main text using slightly modified data from Chua and coworkers and are in good agreement with the $\tau_{surface}$* values obtained in that work.[2] *Predictions of $\tau_{surface}$ from the Random First Order Transition (RFOT) theory and Coupling Model (CM) are plotted as lines.*[1,7]